\documentclass[10pt,superscriptaddress,amsmath,amssymb, aps, pra,twocolumn]{revtex4-2}

\usepackage{mathrsfs}
\usepackage{graphicx}
\usepackage{dcolumn}
\usepackage{bm}
\usepackage[colorlinks]{hyperref}
\usepackage{amsthm}
\usepackage[dvipsnames]{xcolor}
\usepackage{float} 
\usepackage{ragged2e}

\usepackage{xcolor,graphicx}
\usepackage{newlfont}
\usepackage{amssymb,amsmath,mathrsfs}
\usepackage{verbatim}
\usepackage{bbm,braket}
\usepackage{multirow}
\usepackage[varg]{txfonts}
\usepackage{subcaption}
\newcommand{\be}{\begin{equation}}
\newcommand{\ee}{\end{equation}}
\newcommand{\beq}{\begin{eqnarray}}
\newcommand{\eeq}{\end{eqnarray}}

\newcommand{\mc}{\mathcal}

\newcommand{\mbb}{\mathbbm}

\newcommand{\mf}{\mathfrak}

\usepackage{orcidlink}

\usepackage{braket}
  
\begin{document}

\title{Emergence of realism through quantum discord suppression in photonic weak measurements}

\author{Fabr\'{i}cio Lustosa\,}
    \affiliation{Centro de Ci\^{e}ncias Naturais e Humanas, Universidade Federal do ABC-UFABC, Santo Andr\'{e} 09210-580, Brazil}

\author{Diego G. Barreto\,}
    \affiliation{Centro de Ci\^{e}ncias Naturais e Humanas, Universidade Federal do ABC-UFABC, Santo Andr\'{e} 09210-580, Brazil}

\author{Eduardo C. Lima\,}
    \affiliation{Centro de Ci\^{e}ncias Naturais e Humanas, Universidade Federal do ABC-UFABC, Santo Andr\'{e} 09210-580, Brazil}
       
\author{Luciano S. Cruz\,}
     \affiliation{Centro de Ci\^{e}ncias Naturais e Humanas, Universidade Federal do ABC-UFABC, Santo Andr\'{e} 09210-580, Brazil}

 \author{Pedro R. Dieguez\orcidlink{0000-0002-8286-2645}}
\email{pedro.dieguez@ug.edu.pl}
\affiliation{International Centre for Theory of Quantum Technologies, University of Gdańsk, Jana Bażyńskiego 1A, 80-309 Gdańsk, Poland}

\author{Breno Marques\orcidlink{https://orcid.org/0000-0002-2560-7162}}
    \email{breno.marques@ufabc.edu.br}
    \affiliation{Centro de Ci\^{e}ncias Naturais e Humanas, Universidade Federal do ABC-UFABC, Santo Andr\'{e} 09210-580, Brazil}

\begin{abstract}
The emergence of realism from the quantum domain, often associated with the suppression of quantum features, is a key aspect of the quantum-to-classical transition. In this work, we implement an experiment with Werner states subjected to weak measurements to investigate how quantum correlations influence the emergence of realism. Maximally entangled twin photons, generated via spontaneous parametric down-conversion, are used to prepare Werner states. We employ a monitoring model that smoothly transitions between weak and strong nonselective measurements, along with an irrealism measure.  Our findings demonstrate that quantum discord suppression induced by weak measurements, known as weak quantum discord, drives the emergence of realism. Additionally, our findings highlight the robustness of the irrealism measure in quantum correlation-based scenarios.
\end{abstract}

\maketitle

\section{Introduction}
\label{int}

The quantum-to-classical transition, and more broadly the measurement problem, continues to be one of the most intriguing challenges in quantum theory. Understanding how classical phenomena, where observables have well-defined and realistic values, emerge from a quantum framework that lacks intrinsic realism carries deep implications. 
Over the past few decades, foundational debate about the ontic or epistemic nature of wave function has sparked interest~\cite{pusey2012reality,lewis2012distinct,colbeck2012system,hardy2013quantum,patra2013no,aaronson2013psi,leifer2014psi,barrett2014no,ringbauer2015measurements,PhysRevA.75.032110,harrigan2010einstein,hubert2023statistical}. Furthermore, significant conceptual advances in understanding the emergence of objective reality from the quantum substratum have been explored through different approaches like weak measurements~\cite{vaidman1996weak,mancino2018information,dieguez2018information,PhysRevA.111.012220}, decoherence~\cite{schlosshauer2005decoherence}, and quantum darwinism~\cite{zurek2009quantum}.
Monitoring maps, known as generalized measurement maps~\cite{oreshkov2005weak,dieguez2018information,PhysRevA.111.012220,Molitor2024}, which act as a connection between weak and strong (projective) nonselective regimes, have been employed to investigate the emergence of realism from the quantum substratum~\cite{dieguez2018information,mancino2018information,basso2022reality}, and their links with quantum darwinism have been identified~\cite{PhysRevA.111.012220}.  In addition to their foundational importance, weak measurements have proven useful in various applications, such as protecting quantum states from decoherence~\cite{protect1,protect2}, quantum thermal devices~\cite{lisboa2022experimental,dieguez2023thermal,VIEIRA2023100105,malavazi2024weak}, and quantum state tomography~\cite{wu2013state}.

The informational measure known as quantum irrealism~\cite{bilobran2015measure,dieguez2018information,mancino2018information}, derived from the contextual realism hypothesis introduced in~\cite{bilobran2015measure}, is used to assess the degree of realism in quantum systems. The hypothesis generalizes the notion of EPR elements of reality~\cite{PhysRev.47.777}, asserting that in quantum systems, a measured property attains a well-defined value following a projective measurement of a discrete spectrum observable, regardless of whether the specific measurement outcome is known~ \cite{bilobran2015measure,dieguez2018information}. This implies that incoherent mixtures of all possible outcomes have realism for the measured observable.
This measure has proven valuable across a variety of contexts, including coherence~\cite{angelo2015wave}, non-locality~\cite{gomes2018nonanomalous,gomes2019resilience,fucci2019tripartite,gomes2022realism,silva2024does}, random quantum walks~\cite{orthey2019nonlocality}, continuous variables ~\cite{freire2019quantifying,lustosa2020irrealism}, Hardy’s paradox~\cite{engelbert2020hardy}, generalized resource theory of information~\cite{costa2020information}, Aharonov-Bohm effect~\cite{PhysRevA.107.032213}, complementarity relations~\cite{basso2021complete}, delayed-choice arrangements~\cite{dieguez2022experimental}, and quantum eraser experiments~\cite{starke2023correlating}. More recently, generalizations have been proposed, including a theory-independent framework~\cite{fucci2024theory} and a joint reality criterion~\cite{PhysRevA.110.032214}.

In particular, the implications of monitoring for the emergence of realistic properties of quantum two-level systems were experimentally investigated in two separate setups. In the first experiment, a weak measurement was performed in a photonic device~\cite{mancino2018information}, where a fiducial bipartite state was prepared close to a pure state, with initial entropies assumed to be zero within a margin of error comparable to experimental uncertainties~\cite{mancino2018information}. In the second experiment, using IBM's quantum computers, the realism of an observable was examined by monitoring its incompatible counterpart with superconducting qubits~\cite{basso2022reality}. Both experiments corroborated the theoretical prediction that the realism of an observable increases when monitoring the same observable.

In this work, we move forward on the investigation of the emergence of realism by conducting a photonic experiment with Werner state preparation wherein realism induced by the monitoring operation can be completely associated with the amount of quantum discord removed from the measurement procedure. This conclusion is based on a distance-based approach that defines quantum discord in the context of weak measurements, as discussed in~\cite{dieguez2018weak}. The weak quantum discord interpolates between two scenarios, one in which the quantum correlations remain fully intact without measurement and another in which one-way quantum discord vanishes upon performing a local projective measurement~\cite{dieguez2018weak}. This allows weak quantum discord to be understood as a measure of the quantum correlations that are removed by local weak measurements~\cite{dieguez2018weak}.

To implement the experiment, we utilized entangled twin photons created by spontaneous parametric down-conversion (SPDC) to generate a maximally entangled state in polarization \cite{qutoolsquedmanual}. The Werner state and the weak measurement are implemented using the technique discussed in~\cite{lustosa2024simulation}, which is experimentally built through Kraus maps, carrying out unitary operations. 
The manuscript is organized as follows. In Sec.\ref{Sec2}, we introduce the realism quantifier and the weak discord framework. In Sec.\ref{results}, we present our main findings, beginning with the experimental setup used to generate the Werner state and implement the weak measurement. We then demonstrate the emergence of realism through weak quantum discord for various values of measurement strength and initial parameters of the Werner state. Finally, in Sec.~\ref{con}, we summarize our conclusions and offer perspectives for future work.

\section{Theoretical framework}
\label{Sec2}

The realism quantifier is based on the idea that a projective measurement of a discrete-spectrum observable $A=\sum_aaA_a$, with projectors $A_a=\ket{a}\bra{a}$ acting on the Hilbert space $\mc{H_A}$ for a given preparation $\varrho$ on $\mc{H_A\otimes H_B}$ establishes the realism of the measured property related to the observable $A$ even when the measurement outcome is not revealed. Formally, this procedure can be represented by the post-measurement state 
\begin{equation}
 \Phi_A(\varrho):=\sum_a (A_a\otimes\mathbbm{1})\varrho(A_a\otimes \mathbbm{1}),   
\end{equation}
 which is then taken as a primitive notion of $A$-reality state in this framework~\cite{dieguez2018information,dieguez2022experimental}. The quantifier is defined as
\begin{equation}
\label{realism}
\mf{R}_A(\rho):=\ln{d_{\cal{A}}}-\mf{I}_{A}(\rho),
\end{equation}
where 
\begin{equation}
\mf{I}_A(\rho):= \min_{\varrho} S\big(\rho||\Phi_A(\varrho)\big)=S\big(\Phi_A(\rho)\big)-S(\rho)
\end{equation}
is a faithful quantifier of $A$-realism violations for a given state $\rho$ (where $S(\rho||\sigma)=\text{Tr}\big[\rho(\ln{\rho}-\ln{\sigma})\big]$ is the relative entropy, $S(\rho):=\text{Tr}\big(\rho\ln{\rho}\big)$ is the von Neumann entropy) and $d_{\cal{A}}=\dim\mc{H_A}$.
The {\it irrealism} of the  measurement context $A$ and state $\rho$, is bounded as $0\leq\mf{I}_A(\rho)\leq\ln{d_\mc{A}}$ and vanishing iff $\rho=\Phi_A(\rho)$. 

It is evident that whenever quantum discord is present in a bipartite state~\cite{dieguez2022experimental}, the $A$-irrealism induced by the joint state $\rho=\rho_{\mc{A}\mc{B}}$ exceeds that of the subsystem $\mc{A}$ alone. This relation is expressed by the inequality  %
\begin{equation}
\mf{I}_A(\rho)-\mf{I}_A(\rho_\mc{A})\geq \mc{D_A}(\rho),
\end{equation}
where $\rho_\mc{A}=\text{Tr}_\mc{B}(\rho)$ represents the reduced state of subsystem $\mc{A}$. The term $\mc{D_A}(\rho)$ denotes the quantum discord~\cite{Ollivier01,Henderson01,Celeri11,Modi12}, defined as  
\begin{equation}
\mc{D_A}(\rho) := \min_{A} [I_\mc{A:B}(\rho) - I_\mc{A:B}(\Phi_A(\rho))],
\end{equation}  
which quantifies the difference in mutual information before and after a local measurement on $\mc{A}$. Here, mutual information is given by $I_\mc{A:B}(\rho) = S(\rho||\rho_\mc{A} \otimes \rho_\mc{B})$. This result highlights that the presence of quantum discord inherently indicates the $A$-irrealism in the system, connecting the concepts of quantum correlations and the irrealism of local observables.

Another important property concerning maximally incompatible observables $A$ and $A'$ acting on $\mc{H_A}$ 
\be
\label{CP}
\mf{R}_A(\rho)+\mf{R}_{A'}(\rho)\leq\ln{d_\mc{A}}+S(\rho_\mc{A})-I_\mc{A:B}(\rho),
\ee
precludes the manifestation of full realism whenever $\rho\neq \frac{\mbb{1}}{d_A}\otimes\rho_\mc{B}$ and shows that quantum correlations make realism a property that depends not only on the system under investigation but also on any potential correlations it may share~\cite{dieguez2018information,dieguez2022experimental}. For pure states $\rho=\ket{\psi}\bra{\psi}$, the upper bound becomes $\ln{d_\mc{A}}-E(\psi)$, with $E(\psi)=S(\rho_\mc{A(B)})$ the entanglement entropy of $\ket{\psi}$~\cite{dieguez2022experimental}.

\subsection{Weak measurements, realism, and weak quantum discord}

The map referred to as \emph{monitoring},
\be 
M_A^{\epsilon}(\rho):=(1-\epsilon)\rho+\epsilon\,\Phi_A(\rho),
\label{Me}
\ee 
 transitions smoothly between two extremes: no intervention $M_A^{\epsilon \to 0}(\rho) = \rho$ and a projective unrevealed measurement $M_A^{\epsilon \to 1}(\rho) = \Phi_A(\rho)$. For intermediate values $0 < \epsilon < 1$, it represents a weak unrevealed measurement of varying intensity. This map adheres to the non-signaling principle $\text{Tr}_{\mathcal{A}}[M_A^{\epsilon}(\rho)] = \text{Tr}_{\mathcal{A}}[\rho]$ and captures the fact that an infinite sequence of weak measurements converges to a projective measurement $\lim_{n \to \infty}[M_A^{\epsilon}]^n = \Phi_A)$, $\forall~\epsilon\neq0$. Further details on weak maps can be found in Ref.~\cite{dieguez2018information}.

Using the realism quantifier and the hierarchy relation for all measurement strengths $\epsilon$, $M_A^{\epsilon}\Phi_A=\Phi_A M_A^{\epsilon}=\Phi_A$, it was demonstrated in Ref.~\cite{dieguez2018information} that 
\be \label{DR}
\Delta \mathfrak{R}(A)=S(M_A^{\epsilon}(\rho))-S(\rho)\geqslant \epsilon\mf{I}_A(\rho),
\ee 
which is generally non-negative, with equality occurring at $\epsilon = 0$ and $\epsilon = 1$. If $\rho=\Phi_A(\rho)$, then $\Delta\mathfrak{R}=0$, as in this case $\rho$ is already a state of reality for $A$. If $\epsilon\to 1$, 
the variation saturates to the maximum value $\Delta\mathfrak{R}_{\max}(A)=\mf{I}_{A}(\rho)$, meaning that realism maximally emerges when full irrealism is suppressed.
Moreover, one can prove that the following decomposition holds
\be 
\label{LIQCF}
\Delta\mathfrak{R}(A)_{\rho}=\Delta\mathfrak{R}(A)_{\rho_{\cal{A}}}+{D_A}^{\epsilon}(\rho),
\ee  
where  $\Delta\mathfrak{R}(A)_{\rho_{\cal{A}}}=S(M_A^{\epsilon}(\rho_{\cal{A}}))-S(\rho_{\cal{A}})$ is the variation of coherence in the reduced state and ${D_A}^{\epsilon}$ is the non-minimized version of the one-way weak quantum discord~\cite{dieguez2018weak}, defined as
\be 
\cal{D_A}^{\epsilon}(\rho):=\min_A\Big[I(\rho)-I(M_A^{\epsilon}(\rho)) \Big],
\label{De}
\ee 
with $0\leq\epsilon\leq1$. 
The weak quantum discord, as defined by \eqref{De}, has the precise meaning of the amount of quantum discord, presented in the initial state $\rho$, which is destroyed whenever we have a weak non-selective interaction~\cite{dieguez2018weak}.

In the following, we present our experiment to investigate the above decomposition in a specific scenario where the emergence of realism induced by the monitoring map can be associated with the quantum discord suppressed in the process.


\section{Experiment and Results}
\label{results}

\subsection{Experimental setup}
\label{setup}

We used photonic qubits encoded in polarization to implement an experiment showing the validity of the results presented in the previous section. Our experimental configuration includes a twin-photon source generated from spontaneous parametric down-conversion (SPDC). We apply unitary operations to the photon polarization, first to prepare the initial state $\rho$, and later to approximate the action of the monitoring map $M_{A}^{\epsilon}(\rho)$.

\subsubsection{Entangled Photonic States}
\label{setup1}
\begin{figure}
\centering
\includegraphics[width=1\columnwidth]{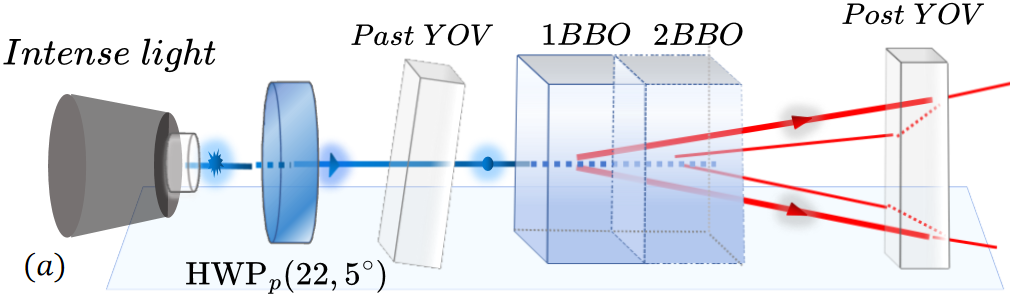}
\includegraphics[width=1\columnwidth]{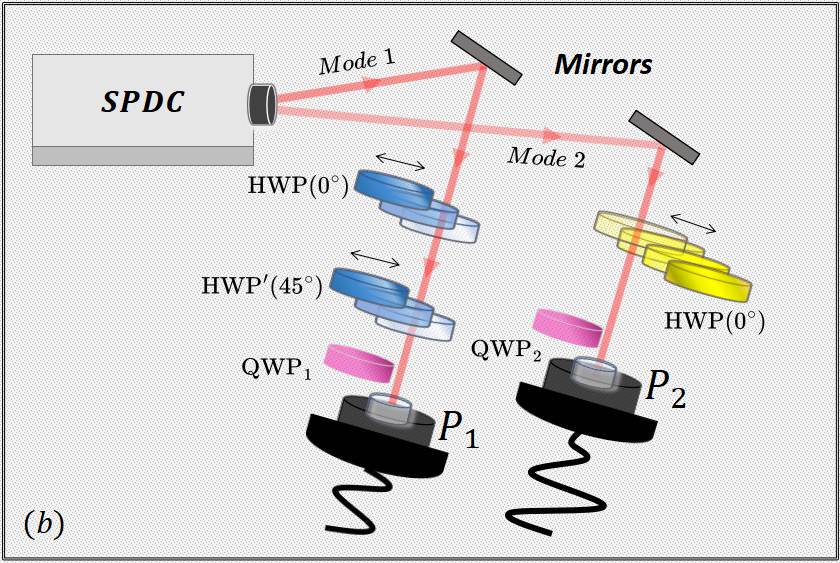}
\caption{(a) Experimental setup for generating the maximally entangled Bell state $|\varphi^{-} \rangle$ in polarization via SPDC in two orthogonal BBO crystal. Twin-photon pair is generated by a pump laser at diagonal polarization, achieved using $HWP_p$ set at $22.5^\circ$, enabling down-conversion in either the first or second crystal. (b) In mode 1, a combination of two HWPs is used to prepare the Werner state, as detailed in subsection~\ref{setup2}. In mode 2, the monitoring map is implemented inserting or not a HWP, as described in subsection~\ref{WMI_subsec}. Subsequently, QWPs and Polarizers are employed to make the measurement for quantum state tomography. The twin photons are coupled into optical fibers and directed to single-photon detectors for coincidence counts.\justifying}
\label{spdc1}
\end{figure}

The schematic of our experimental setup is shown in Figure~\ref{spdc1}. In the first part, a diode laser of $50$~mW emits a beam at a wavelength of $405$~nm and pumps two $\beta$ -barium boreate (BBO) crystals, which have been cut adequately for carrying out collinear type I phase matching, and the optical axes are held orthogonal to each other, as illustrated in Figure~\ref{spdc1}(a). In this way, a half-wave plate (HWP) set at $22.5^{\circ}$, fixed before the non-linear crystals, turns the pump photon polarization $|H \rangle, |V \rangle$ into $\frac{1}{\sqrt{2}}( |H \rangle \pm  |V \rangle )$, respectively. Additionally, a pair of compensating Yttrium vanadate (YVO) crystals is placed to correct temporal and spatial effects that reduce the coherence of the system \cite{qutoolsquedmanual}. Accordingly, the pair of down-conversion photons from the coupled BBO crystals is indistinguishable, and the initial state of the system is prepared to approximate a Bell state in polarization, $|\varphi^{-}\rangle = \frac{1}{\sqrt{2}}[|H_1 \rangle |H_2 \rangle - |V_1 \rangle |V_2 \rangle]$. 

In Figure~\ref{spdc1} (b), the down-converted photons are reflected by the mirrors, defining specified single spatial modes. Photon polarization analysis is carried out using adjustable quarter-wave plates, QWP$_{1(2)}$, plus polarizers, $P_{1(2)}$, allowing us to reconstruct a quantum state by tomography measurement (combination of all Pauli matrices). Each mode is equipped with an optical fiber coupler that steers the photons to be detected in coincidence by avalanche single-photon detectors. 

\subsubsection{Preparation of Werner States}
\label{setup2}
The SPDC process allows the preparation of Werner states by submitting one of the photons to a controlled depolarization channel of one qubit~\cite{lustosa2024simulation, zhang2002experimental, liu2017experimental}. It was recently demonstrated in~\cite{lustosa2024simulation} that depolarizing dynamics can be achieved by dividing the acquisition time into intervals, each corresponding to different operations.
In this case, the Pauli operators ($\sigma_x$, $\sigma_y$, $\sigma_z$) contribute equally to the dynamics, while the identity operator $\mbb{1}$ plays a distinct role with a different weight.

The operators needed to prepare the Werner state are 
\begin{equation}
    K_0 = \sqrt{p_0}\,\mbb{1}, \;\;\; K_x = \sqrt{p_x}\sigma_x \nonumber
\end{equation}
\begin{equation}
    K_y =  \sqrt{p_y}\sigma_y,\, \;\; K_z = \sqrt{p_z}\sigma_z,
 \label{operators}  
\end{equation}
where $p_0 + p_x + p_y + p_z = 1$ guarantees the property $\sum_{k}K_{\mu}^{\dagger}K_{k}=\mbb{1}$. Figure \ref{spdc1}(b) shows how to implement operators $K_i$ using the wave plates in mode 1. We insert the plates $\text{HWP}(0)$ and $\text{HWP}^{\prime}(45^\circ)$ in the path through which the photons from mode 1 pass through. The Pauli operators are run by rotating the photon polarization. For example, the angles required for each operation produce $\sigma_x = \text{HWP}(45^\circ)$, $\sigma_y =  \text{ HWP}(0^\circ) \text{HWP}^{\prime}(45^\circ)$, and $\sigma_z =  \text{ HWP}^{\prime}(0^\circ)$ \cite{lustosa2024simulation}.  When we apply these transformations, we get the four maximally entangled Bell states:
\begin{equation}
|\varphi^{-} \rangle = \Bigg( \frac{K_0}{\sqrt{p_{0}}} \otimes \mbb{1}\Bigg)|\varphi^{-} \rangle, \; |\psi^{-} \rangle  = \Bigg( \frac{K_x}{\sqrt{p_{x}}} \otimes \mbb{1}\Bigg)|\varphi^{-} \rangle \nonumber
\end{equation}
\begin{equation}
\;\;\;\;\;\;|\psi^{+} \rangle =  \Bigg( \frac{K_y}{\sqrt{p_{y}}} \otimes \mbb{1} \Bigg)|\varphi^{-} \rangle, \; |\varphi^{+} \rangle  = \Bigg( \frac{K_z}{\sqrt{p_{z}}} \otimes \mbb{1}\Bigg)|\varphi^{-} \rangle,
 \label{Bell states}
\end{equation}
where the normalization is omitted in this equation.  

Considering the effect of map \eqref{operators} in the source state $\rho = \ket{\varphi^{-}}\bra{\varphi^{-}}$, we obtain  
\begin{equation}
    \mathcal{E}(|\varphi^{-} \rangle \langle \varphi^{-}|) =
\sum_{i=0}^z M_i |\varphi^{-} \rangle \langle \varphi^{-}| M_{i}^{\dagger}, \nonumber
\end{equation}
\begin{equation}
    \mathcal{E}(|\varphi^{-} \rangle \langle \varphi^{-}|) =
p_0\ket{\varphi^{-}}\bra{\varphi^{-}} + p_x\ket{\psi^{-}}\bra{\psi^{-}} \nonumber
\end{equation}
\begin{equation}
\;\;\;\;\;\;\;\;\;\;\;\;\;\;\;\;\;\;\;\; + p_y\ket{\psi^{+}}\bra{\psi^{+}}  + p_z\ket{\varphi^{+}}\bra{\varphi^{+}}.
\end{equation}
Given that the parameters $p_i$ fulfill $p_0 \rightarrow (1+3\mu)/4; \; p_x=p_y=p_z \rightarrow (1-\mu)/4$, the state is represented by
\begin{equation}
|\varphi^{-} \rangle \langle \varphi^{-}| \longmapsto (1 - \mu)\frac{\mathbbm{1} \otimes \mathbbm{1}}{4} + \mu|\varphi^{-} \rangle \langle \varphi^{-}|.
\end{equation}

To implement the state, we divided the acquisition time $\Delta T=16s$ between the four operations with the given subdivision time $\vec{t}_w^{~T}=(\Delta t_0,\Delta t_x,\Delta t_y, \Delta t_z)$. The effect of the depolarization channel occurs at intervals ranging from $0$ to $\Delta T$ where $\mu = \Delta t_i / \Delta T$  given the preparation $\rho_\mu$ implemented in this work listed below.

\begin{align}
\mu &= 1, \quad \vec{t}_w = \begin{pmatrix}  16s\\0\\0\\0  
\end{pmatrix} \; \rightarrow \; \rho_1 = |\varphi^{-}\rangle \langle \varphi^{-}|, 
\\
\mu &= \frac{3}{4}, \quad \vec{t}_w = \begin{pmatrix} 13s \\ 1s \\ 1s \\ 1s \end{pmatrix} \; \rightarrow \; \rho_{3/4} = \frac{3}{4} |\varphi^{-}\rangle \langle \varphi^{-}| + \frac{1}{16} \; \mbb{1} \otimes \mbb{1}, 
\\
\mu &= \frac{1}{2}, \quad \vec{t}_w = \begin{pmatrix} 10s \\ 2s \\ 2s \\ 2s \end{pmatrix} \; \rightarrow \; \rho_{1/2} = \frac{1}{2} |\varphi^{-}\rangle \langle \varphi^{-}| + \frac{1}{8} \; \mbb{1} \otimes \mbb{1}, 
\\
\mu &= \frac{1}{4}, \quad \vec{t}_w = \begin{pmatrix} 7s \\ 3s \\ 3s \\ 3s \end{pmatrix} \; \rightarrow \; \rho_{1/4} = \frac{1}{4} |\varphi^{-}\rangle \langle \varphi^{-}| + \frac{3}{16} \; \mbb{1} \otimes \mbb{1}, 
\\
\mu &= 0, \quad \vec{t}_w = \begin{pmatrix} 4s \\ 4s \\ 4s \\ 4s \end{pmatrix} \; \rightarrow \; \rho_0 = \frac{1}{4} \mbb{1} \otimes \mbb{1}.
\end{align}
 Initially, the physical system is in $|\varphi^{-}\rangle \langle \varphi^{-}|$, and throughout the acquisition time, only $K_0$ operates for the entire duration $\Delta T$. However, as the operating time of the operators $K_i$ increases, they become a significant part of the acquisition time. This gradually introduces errors until the system reaches its most degraded state at $\Delta t_i = \Delta T = 16s$, $i=\{0,x,y,z\}$.  At this point, the system state is maximally mixed.

\subsubsection{Weak Measurement Implementation}
\label{WMI_subsec}
In equation \ref{Me}, we defined the monitoring map $M_A^{\epsilon}$ given the strength $\epsilon$ and the observable $A$. Considering the observable with projectors $A_{0}=\ket{0}\bra{0}$ and $A_{1}=\ket{1}\bra{1}$, we can rewrite this monitoring maps act in $\rho$ as:
\begin{eqnarray}
    M_A^\epsilon(\rho)&=&(1-\epsilon)\rho+\epsilon\left[\bra{0}\rho\ket{0}+\bra{1}\rho\ket{1}\right].
\end{eqnarray}
Note that, the application of the monitoring map gives the same result as the dephasing map
\begin{equation}
     M_A^\epsilon(\rho)=\left(1-\frac{\epsilon}{2}\right)\rho+\frac{\epsilon}{2}\sigma_z\rho\sigma_z.
\end{equation}

We applied the same technique used to prepare the Werner state, we can use it to implement the monitoring map for an observable that has the computational basis as a spectrum. For our experiment, we used the acquisition time $\Delta T =16s$ and the subdivision $\vec{t}_d^{~T}=(t_0,t_z)$, where $t_0$ ($t_z$) is the time interval that is applied identity ($\sigma_z$). In this work, we implement the following monitoring maps:

\begin{align}
\epsilon &= 0, \quad \vec{t}_w = \begin{pmatrix}  16s\\0s
\end{pmatrix} \; \rightarrow \; M_A^{0}(\rho)=\rho,
\\
\epsilon &= \frac{1}{4}, \quad \vec{t}_d = \begin{pmatrix} 14s \\ 2s \end{pmatrix} \; \rightarrow \; M_A^{1/4}(\rho)=\frac{3}{4}\rho+\frac{1}{4}\sum_i\bra{i}\rho\ket{i}, 
\\
\epsilon &= \frac{1}{2}, \quad \vec{t}_d = \begin{pmatrix} 12s \\ 4s \end{pmatrix} \; \rightarrow \; M_A^{1/2}(\rho)=\frac{1}{2}\rho+\frac{1}{2}\sum_{i}\bra{i}\rho\ket{i},
\\
\epsilon &= \frac{3}{4}, \quad \vec{t}_d = \begin{pmatrix} 10s \\ 6s \end{pmatrix} \; \rightarrow \;M_A^{3/4}(\rho)=\frac{1}{4}\rho+\frac{3}{4}\sum_i\bra{i}\rho\ket{i},
\\
\epsilon &= 1, \quad \vec{t}_d = \begin{pmatrix} 8s \\ 8s  \end{pmatrix} \; \rightarrow \; M_A^{1}(\rho)=\sum_i\bra{i}\rho\ket{i}.
\end{align}

\begin{figure}[h]
    \centering
    \includegraphics[width=1.0\columnwidth]{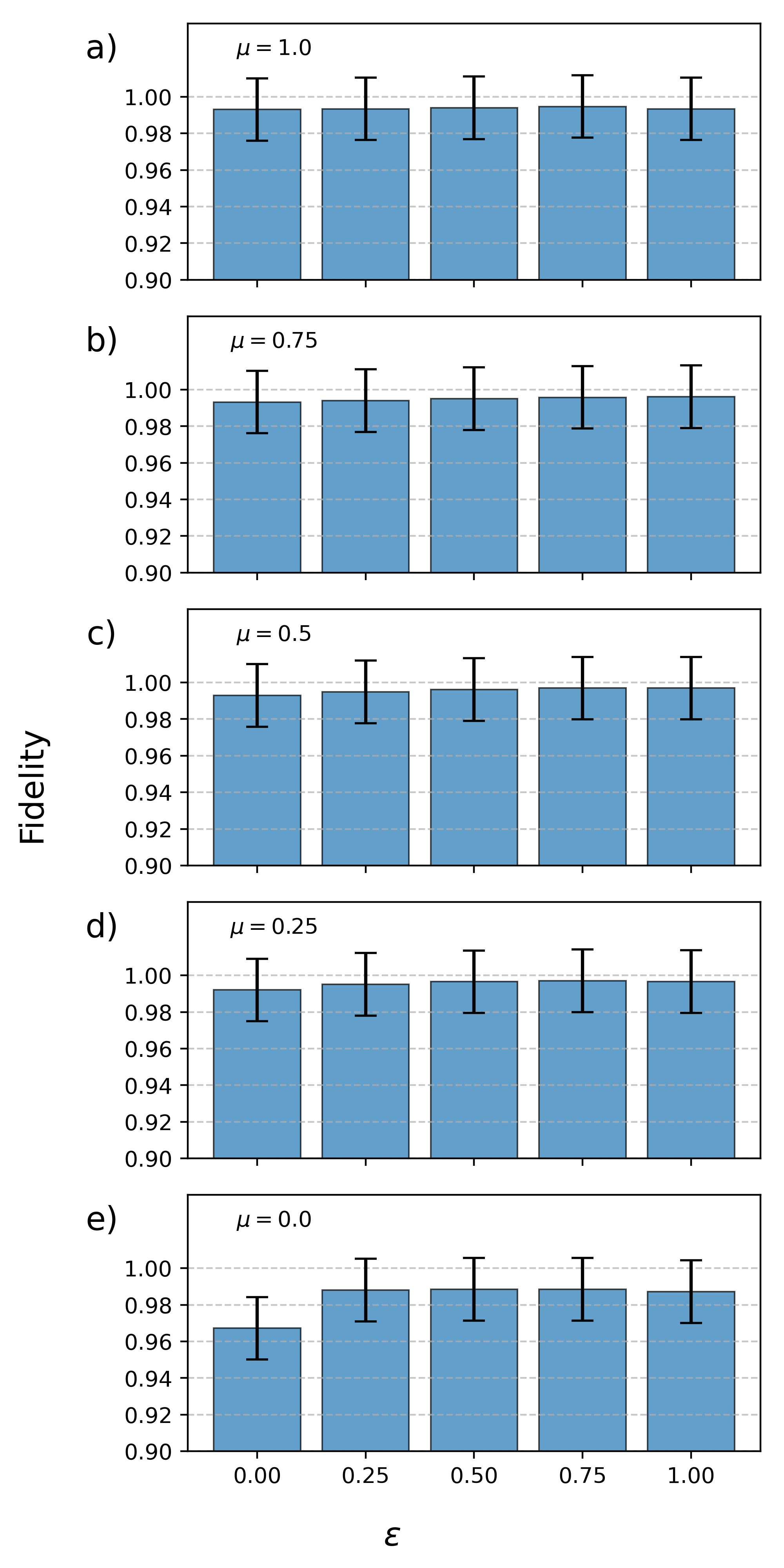}
    \caption{Fidelity of the implemented states as a function of the measurement strength. Subfigures (a), (b), (c), (d), and (e) correspond to $\mu=1$, $\mu=3/4$, $\mu=1/2$, $\mu=1/4$, and $\mu=0$, respectively. The fidelity is calculated between the ideal case with the initial state as $\ket{\varphi^{-}}$ and the measure state for all implemented values of $\mu$ and $\epsilon$.\justifying}
    \label{fidelidade}
\end{figure}

In Figure~\ref{spdc1}(b), mode 1 is used to prepare the Werner state and, in mode 2, we implement the monitoring map, using an HWP at $0^\circ$ to implement $\sigma_z$ and removing it to implement the identity. To calculate the weak quantum discord properties of the state for each Werner state and monitoring map strength, we performed quantum tomography. Taking the initial state as $\ket{\varphi^{-}}$, we calculate the fidelity for each $\mu$ and $\epsilon$, which can be seen in Figure~\ref{fidelidade}.

\subsection{Emergence of realism via weak quantum discord}

\begin{figure}
\centering
\includegraphics[width=1.0\columnwidth]{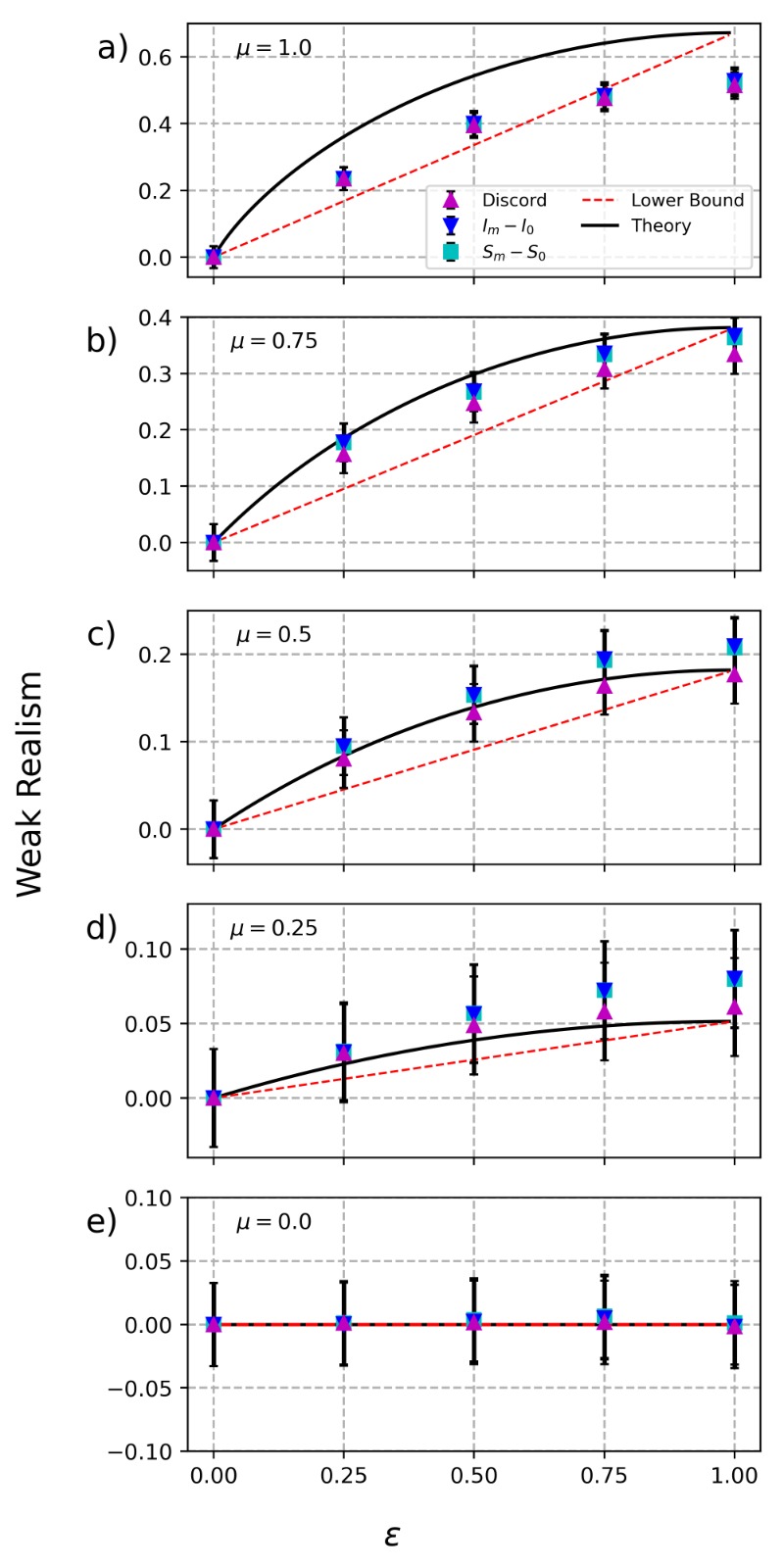}
\caption{Weak realism as a function of the measurement strength. The results demonstrate that tuning the measurement strength increases the realism for different values of $\mu$ in the Werner state. Specifically, subfigures (a), (b), (c), (d), and (e) correspond to $\mu=1$, $\mu=3/4$, $\mu=1/2$, $\mu=1/4$, and $\mu=0$, respectively. Considering the ideal scenario, with the $\ket{\varphi^{-}}$ as the source state, the solid black curve represents the theoretical prediction while the dashed red curve corresponds to the lower bound for the amount of weak realism obtained via monitoring. The experimental results are depicted using up-and-down triangles and squares, each accompanied by its respective error bars, representing respectively, the experimental minimization used to determine quantum discord, the expression based on the difference in mutual information, and the realism variation. \justifying}
\label{gráficos}
\end{figure}

 For the Werner state
 \begin{equation}
     \rho^{\mu}=(1 - \mu)\frac{\mathbbm{1} \otimes \mathbbm{1}}{4} + \mu|\varphi^{-} \rangle \langle \varphi^{-}|,
 \end{equation}
 the local state $\rho^{\mu}_{\cal{A}}=\text{Tr}_{\mathcal{B}}[\rho_{\cal{AB}}]$ is not affected by the monitoring since $\rho^{\mu}_{\cal{A}}=M_A^{\epsilon}(\rho^{\mu}_{\cal{A}})=\mbb{1}/2$.
 One can use the above condition, the invariance under local rotations, and the lower bound given by Eq.~\ref{DR} to write 
 \be 
\Delta\mathfrak{R}(A)_{\rho^{\mu}}=\cal{D_A}^{\epsilon}(\rho^{\mu})=S(M_A^{\epsilon}(\rho^{\mu}))-S(\rho^{\mu})\geqslant \epsilon\mf{I}_A(\rho^{\mu})
\label{main}.
\ee  
 The equality reveals the connection between the emergence of realism and weak quantum discord for Werner states. It holds by evaluating the eigenvalues of $M_A^{\epsilon}(\rho^{\mu})$ with a generic observable $A(\theta,\phi)=\sum_{a=\pm} a A_a(\theta,\phi)$ with projectors $A_{\pm}(\theta,\phi)=\ket{\pm}\bra{\pm}$, where the eigenstates are given by $\ket{+} =\cos{\left(\tfrac{\theta}{2}\right)}\ket{0} +e^{i\phi}\sin{\left(\tfrac{\theta }{2}\right)}\ket{1}$ and $\ket{-}=\sin{\left(\tfrac{\theta}{2}\right)}\ket{0} -e^{i\phi}\cos{\left(\tfrac{\theta}{2}\right)}\ket{1}$. The transformation of the state under measurement is given by $\Phi_{A(\theta,\phi)}(\rho^{\mu})=\sum_{a=\pm}(A_a(\theta,\phi)\otimes \mbb{1})\rho^{\mu}(A_a(\theta,\phi) \otimes \mbb{1} )$, leading to the weakly measured state $M_{A(\theta,\phi)}^{\epsilon}(\rho^{\mu})=(1-\epsilon)\rho^{\mu}+\epsilon\Phi_A(\rho^{\mu})$. The eigenvalues of this state are found to be $\lambda_1 = \tfrac{1-\mu}{4}$, $\lambda_2 = \tfrac{1-\mu}{4}$, $\lambda_3 = \tfrac{1+3\mu-2\mu\epsilon}{4}$, and $\lambda_4 = \tfrac{1-\mu+2\mu\epsilon}{4}$. Importantly, these eigenvalues do not depend on the parameters $(\theta, \phi)$, reflecting the rotational invariance of the Werner state. This allows us to choose any basis for $A$, including the computational basis, the one implemented in this experiment.
The result then reads
\be
\Delta\mathfrak{R}(A)_{\rho^{\mu}}=\tfrac{1}{4}\sum_{i=-1}^1\sum_{j=0}^1(-1)^j\lambda_{ij}\ln{\lambda_{ij}},
\label{Derhomu}
\ee 
with $\lambda_{ij}=1+\mu[1+2i(1-j\epsilon)]$.

Figure~\ref{gráficos} presents our main results, illustrating how weak realism emerges from the suppression of quantum discord induced by local measurements that transition smoothly between weak and strong regimes. Furthermore, we examine the robustness of the lower bound in Eq.~\ref{main}, extending the analysis to a scenario involving quantum correlations, going beyond previous studies that focused on quantum coherence in pure states~\cite{mancino2018information}. The findings emphasize that, for Werner states, the emergence of weak realism is directly linked to the suppression of quantum discord induced by weak measurements, a phenomenon known as weak quantum discord.

\section{Discussion}
\label{con}

One key advantage of the quantum irrealism measure is its quantitative and operational nature, aligning with the EPR criterion by ensuring that eigenstate preparations are always elements of reality for some observable. In addition, it extends the EPR concept by quantifying the degree of realism for mixed states. This approach highlights the fundamental role of information in the quantum-to-classical transition~\cite{dieguez2018information}. Realism emerges as quantum features are suppressed, whether through the loss of quantum coherence in a given basis or the reduction of quantum correlations such as quantum discord.

Previous experiments investigated the emergence of realism using single-partite coherent states~\cite{mancino2018information}, where the transition to realism was driven by the suppression of quantum coherence. 
In contrast, our experiment explores a different aspect of this phenomenon by focusing on quantum correlations rather than coherence. Specifically, we analyze how tuning the measurement strength suppresses quantum discord, leading to the emergence of realism.  Although the results obtained with a maximally entangled state exhibited some noise in the strong measurement regime given by our twin-photon source, the remaining data closely aligned with theoretical predictions and respected the proposed lower bound. This demonstrates the robustness of our approach and further supports the role of quantum discord suppression in the emergence of classical realism, highlighting the feasibility of the irrealism measure in a regime governed by quantum correlations.

To probe this relationship, we performed an experiment using maximally entangled twin photons generated through spontaneous parametric down-conversion to prepare the Werner state, followed by photonic weak measurements. Werner states provide a well-controlled platform for studying quantum correlations, allowing us to tune entanglement and quantum discord. By employing monitoring maps~\cite{dieguez2018information}, we systematically examined how suppressing quantum discord leads to the emergence of realism. Our experiment further demonstrates that weak measurements are effective tools for probing the quantum-to-classical transition, enabling a controlled shift between weak and projective measurements.
The results confirm this connection, as shown by data represented with up-and-down triangles and squares, each with corresponding error bars. These markers correspond to three distinct methods of analysis: the experimental minimization used to determine quantum discord, the expression based on the difference in mutual information, and the measure of realism obtained through the monitoring procedure. The consistency across these approaches reinforces the validity of our findings, providing evidence that weak measurements performed in Werner states progressively suppress quantum discord, driving the emergence of classical realism. 

Building on this work, future advancements of our setup could be utilized to explore the recently proposed connection between the irrealism measure and complete complementarity relations~\cite{basso2021complete}, offering new insights into the interplay between quantum correlations~\cite{dieguez2018weak}, realism~\cite{bilobran2015measure,dieguez2018information}, and complementarity~\cite{dieguez2022experimental,spegel2024experimental,PhysRevA.109.062207}.

\section*{Acknowledgements}
F.L. acknowledges Coordenação de Aperfeiçoamento de Pessoal de Nível Superior (CAPES). D.G.B and E.C.L. acknowledge support from Fundação de Amparo a Pesquisa do Estado de São Paulo (FAPESP) (Grant No. 2023/03513-0  and No. 2023/11444-9). P.R.D. acknowledges support from the NCN Poland, ChistEra-2023/05/Y/ST2/00005 under the project Modern Device Independent Cryptography (MoDIC). B.M. acknowledges the Conselho Nacional de Desenvolvimento e Tecnológico (CNPq). All the authors acknowledge support from Instituto Nacional de Ciência e Tecnologia de Informação Quântica (CNPq-INCT-IQ Grant No. 465469/2014-0) and FAPESP (No. 2021/14303-1). 

\bibliography{citations}

\end{document}